\documentclass[english,aps,superscriptaddress,twocolumn]{revtex4}

\usepackage[T1]{fontenc}

\usepackage{graphicx}
\usepackage{epstopdf}
\usepackage{amssymb}
\usepackage{amsmath}
\usepackage{lipsum}
\usepackage{subfigure}
\usepackage[urlcolor=blue,hyperindex,colorlinks,bookmarks=true,linkcolor=black,citecolor=black]{hyperref}
\usepackage[normalem]{ulem}
\usepackage[abs]{overpic}
\usepackage{color}
\usepackage{rotating}

 \newcommand{\be}{\begin{equation}}
\newcommand{\ee}{\end{equation}}
\newcommand{\bea}{\begin{eqnarray}}
\newcommand{\eea}{\end{eqnarray}}

\newcommand{\ket}[1]{\left|#1\right\rangle}
\newcommand{\bra}[1]{\left\langle#1\right|}

\newcommand{\abs}[1]{\lvert#1\rvert}

\newcommand{\al}{\alpha}

\begin{document}
\title{Generating Nonclassical States from Classical Radiation by Subtraction Measurements}

\author{Luke C.G. Govia}
\affiliation{Theoretical Physics, Universit\"{a}t des Saarlandes, Saarbr\"{u}cken, Germany}
\affiliation{Institute for Quantum Computing and Department of Physics and Astronomy, University of Waterloo, Ontario, Canada}
\author{Emily J. Pritchett}
\affiliation{Theoretical Physics, Universit\"{a}t des Saarlandes, Saarbr\"{u}cken, Germany}
\author{Frank K. Wilhelm}
\affiliation{Theoretical Physics, Universit\"{a}t des Saarlandes, Saarbr\"{u}cken, Germany}
\affiliation{Institute for Quantum Computing and Department of Physics and Astronomy, University of Waterloo, Ontario, Canada}

\begin{abstract}
We describe the creation of nonclassical states of microwave radiation via ideal dichotomic single photon detection, i.e., a detector that only indicates presence or absence of photons. Ideally, such a detector has a back action in the form of the subtraction operator.  Using the non-linearity of this back action, it is possible to create a large family of nonclassical states of microwave radiation, including squeezed and multi-component cat states, starting from a coherent state. We discuss the applicability of this protocol to current experimental designs of Josephson Photomultipliers (JPMs).
\end{abstract}
\maketitle

\section{Introduction}
The generation of nonclassical states of radiation is an important test of the foundations of quantum mechanics and a necessary precursor to implementing quantum communication and computation protocols in many architectures \cite{Knill01,Braunstein05,Gisin02}.  While the methodology for creating nonclassical radiation at optical wavelengths has been studied extensively \cite{Lvovsky:2009fk,Ourjoumtsev:2006uq,Shi-Biao:2005fk,Dakna:1997fk}, the technology to create quantum states with larger and larger wavelengths has recently become available with advances in cavity- and circuit-QED.

In this paper we present a novel way to generate a family of nonclassical states of microwave radiation in a long wavelength transmission line using only detection by an ideal binary detector, such as the Josephson Photomultiplier (JPM). The protocol only involves radiating a microwave cavity with coherent radiation and post selection based on single photon detection, without further manipulation.
In addition, our protocol applies to any detection mechanism with a back action resembling that of the subtraction operator (equation (\ref{eqn:Sub})) and so can be generalized to other quantum systems, in particular, other superconducting circuits where strong photon-detector coupling is possible \cite{Akhlaghi:2011mz,Hofheinz:2008ly,Hofheinz:2009qf}. Recent proposals have established an analogous detection scheme in cavity-QED, broadening the range of application of our results \cite{Raimond:2010fk,Oi:2013fk}.

In the microwave regime of cavity-QED/circuit-QED squeezed states \cite{Castellanos-Beltran:2008vn,Eichler:2011ys,Wilson11} and cat-like states \cite{Kirchmair:2013uq,Deleglise:2008fk}  of microwave radiation have been generated by the Kerr interaction between a cavity/transmission line and coupled atoms/superconducting qubits. Multi-component cat states have also been produced in circuit-QED using a gate-based construction \cite{Hofheinz:2008ly,Hofheinz:2009qf}.   We show how these nonclassical states can be created in circuit-QED by a measurement based protocol, and add a new class of nonclassical states to the list, the generalized squeezed states, which so far have only been proposed in theory \cite{Braunstein:1987kx,Braunstein:1990vn}. 

The JPM, a current biased Josephson junction related to the phase qubit, has been shown experimentally \cite{Chen:2011fk} and theoretically \cite{Govia:2012uq,Poudel:2012uq,Peropadre11} to be an effective single microwave photon counter.  Previously, we have shown that for a JPM under optimal conditions the back action of photon detection is the photon subtraction operator \cite{Govia:2012uq},
\begin{equation}
\hat{B}\equiv\sum_{n=1}^\infty|n-1\big>\big<n|,
\label{eqn:Sub}
\end{equation}
a nonlinear operator that can be related to the photon lowering operator by $\hat{a} = \hat{B}\sqrt{\hat{n}}$, but cannot be expressed as a linear combination of photon creation and annihilation operators. Also, note that $\hat{B}$ is not invertible, and hence not unitary. The JPM can be seen in this regime as an ideal dichtomic detector, providing information about the presence or absence of photons but not revealing their number beyond that. 

\section{Protocols for Nonclassical State Generation}

In this paper we show how to use the noncommuntativity of the detection back action with coherent displacement pulses to achieve single mode quadrature squeezing of an input coherent state as well as to generate other nonclassical states, namely generalized squeezed states and squeezed multi-component Schr\"{o}dinger cat states \cite{Lvovsky:2009fk}. Note that a special case of the second step of our protocol is already known in quantum optics: subtracting a photon from a squeezed vacuum state produces a low-power cat (kitten) state \cite{Dakna:1997fk}.

\subsection{Squeezed States}
The generation of squeezed states of microwave radiation using JPMs follows a simple protocol. The cavity is initially prepared in a coherent state, $|\alpha\big>=\hat{D}(\alpha)|0\rangle=e^{-\frac{|\alpha|^2}{2}}\sum_{n=0}^\infty\frac{\alpha^n}{\sqrt{n!}}|n\big>$ where $\alpha\equiv |\alpha|e^{i\varphi_\alpha}$, and is coupled to one or more detectors, each acting with back action $\hat{B}$ on the cavity after a photon is detected. Mathematically a coherent displacement such as this is represented by the displacement operator $\hat{D}(\al) = e^{\left(\al\hat{a}^{\dagger} - \al^{*}\hat{a}\right)}$.  After $N$ photons are counted, a further displacement pulse is applied such that the state is centred around  $-\alpha$ in phase space. After $N$ further photon detections are observed, the resulting state is a squeezed state. The optimal choice of $N$ will be discussed shortly.

Starting from the coherent state input, the probability for $N$ photons to be detected is 
\begin{equation}
\label{eqn:ProbN}
P_N\equiv 1-e^{-|\alpha|^2}\sum_{n=0}^{N-1}\frac{|\alpha|^{2n}}{n!} = 1 - \frac{\Gamma(N,\abs{\al}^2)}{\Gamma(N)}
\end{equation}
where $\Gamma(N,\abs{\al}^2)$ is the upper incomplete gamma function of $N$ and $\abs{\al}^2$ \cite{George-B.-Arfken:2012fk}. $P_N\sim|\alpha|^{2N}/N!$ as $|\alpha|\rightarrow 0$; however, at $\abs{\al}^2 \approx N$ $P_N$ jumps rapidly towards unity, and so can be made arbitrarily close to unity with higher power coherent pulses (see appendix \ref{ap:Prob} for further detail). 

It is straightforward to calculate the normalized post measurement cavity state after $N$ detections, 
\begin{equation}
\rho'\equiv\frac{\hat{B}^N|\alpha\big>\big<\alpha|\hat{B}^{\dagger  N}}{P_N},
\end{equation}
and the average photon number $n_1 \equiv\big<a^\dagger a\big>_{\rho '}$ is given by
\bea
n_1 = \frac{\abs{\alpha}^2\left( 1 - \frac{\Gamma(N-1,\abs{\al}^2)}{\Gamma(N-1)} \right) - N\left( 1 -  \frac{\Gamma(N,\abs{\al}^2)}{\Gamma(N)} \right)}{P_N},
\eea
which can also be numerically evaluated. After $N$ detections, the next step is to displace the state by an amount $\alpha_1=-\sqrt{n_1}e^{i\varphi_\alpha} -\alpha$, so that the resulting state will be centred in phase space around $-\al$. For this displaced state, $N$ photon detection events will occur with probability $P_N'$
 \footnote{for which the analytic expression is cumbersome and not instructive but has the same limiting values as $P_N$ for $\abs{\al} \rightarrow \infty$}
, and the renormalized cavity state will have the form
\begin{equation}
\label{eqn:SqueezedState}
\rho''\equiv\frac{\hat{B}^ND(\alpha')\rho'D(\alpha')^\dagger\hat{B}^{\dagger N}}{P_N'}.
\end{equation}
We will now show that the state $\rho''$ is a squeezed state. 

\begin{figure}[h!]
\subfigure{
\label{fig:DeltaP}
\begin{overpic}[width=\columnwidth]
{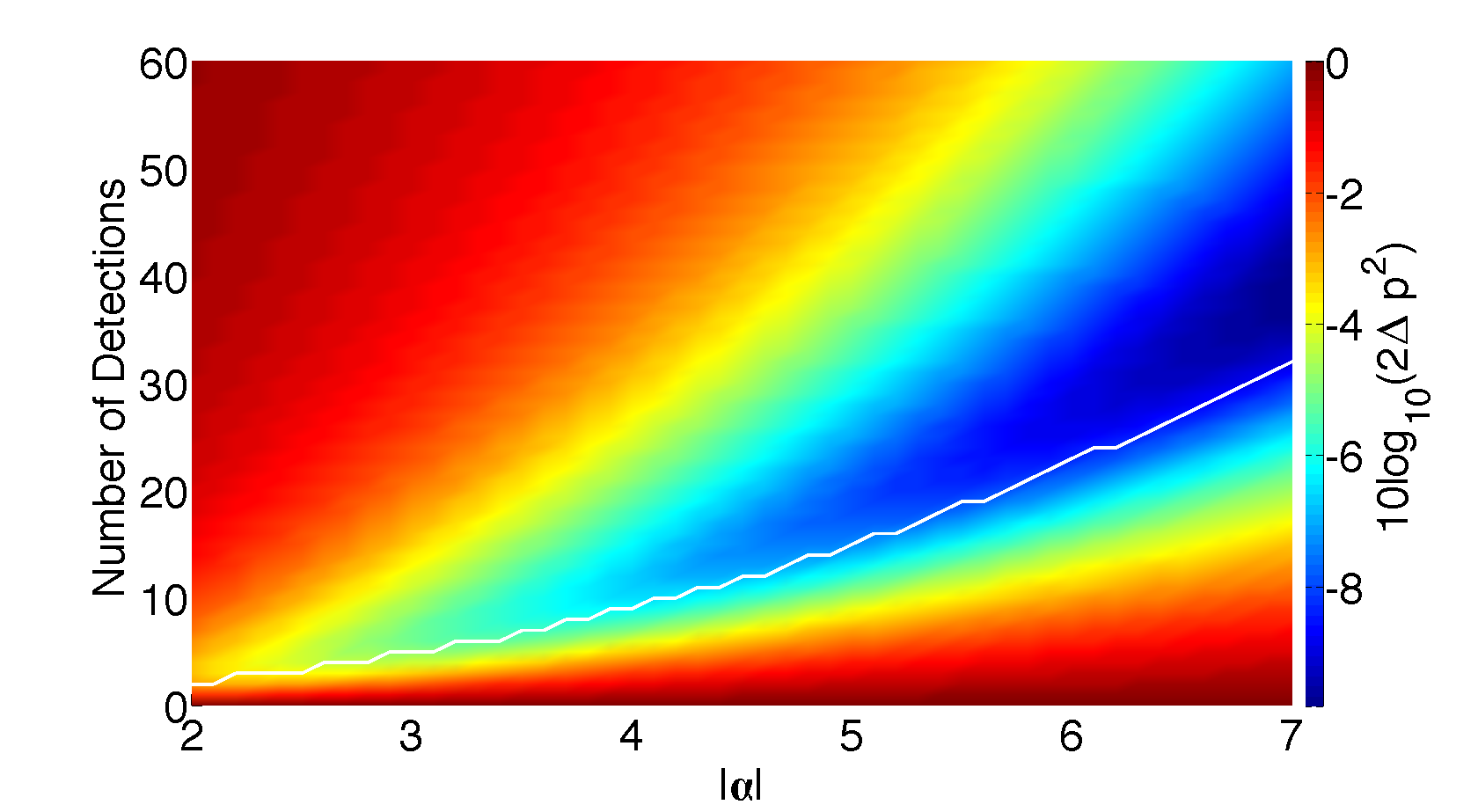}
\put(40,110){{\color{white} \bf (a)}}
\end{overpic}}
\subfigure{
\label{fig:HB}
\begin{overpic}[width=\columnwidth]
{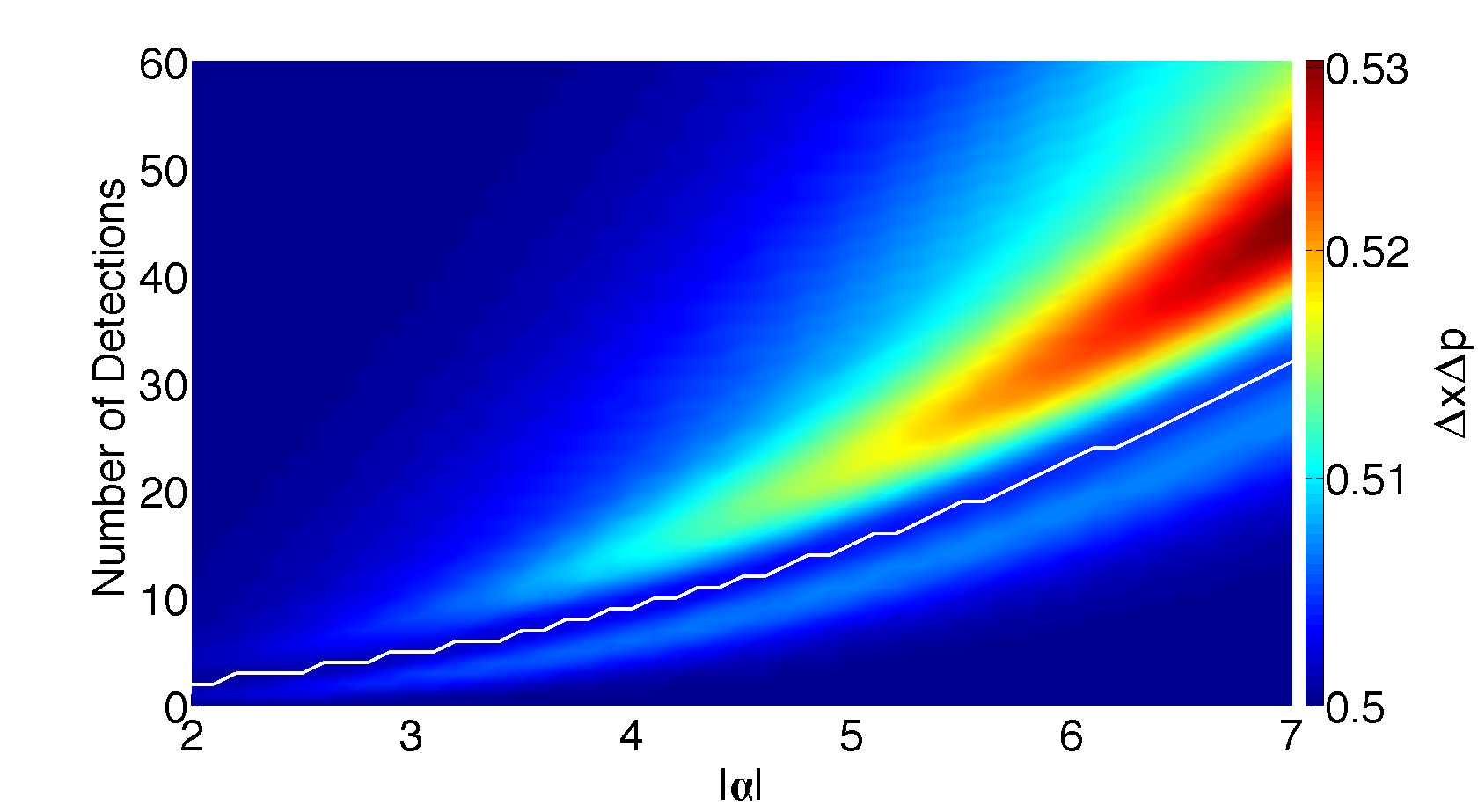}
\put(40,110){{\color{white} \bf (b)}}
\end{overpic}}
\caption{(a) $S(\Delta p)$ and (b) $\Delta x \Delta p$ of the state $\rho''$, as functions of $\al$ and the number of detection events, $N$. $\varphi_{\al} = 0$ for both figures. The white curve indicates the value of $N$ that gives a local minimum in $\Delta x \Delta p$, while maintaining a squeezed $\Delta p^2$. }
\end{figure}

To quantify the amount of squeezing, we calculate the variance of the squeezed quadrature
\be
\Delta p^2 = {\rm Tr}[\rho'' \hat{p}^2] - {\rm Tr}[\rho'' \hat{p}]^2.
\ee
The quadrature observable $\hat{p}(\varphi_a)$ is defined by $\hat{p}(\varphi_a) = \frac{1}{\sqrt{2}}\left( \hat{a}e^{-i\varphi_\alpha} - \hat{a}^{\dagger}e^{i\varphi_\alpha}\right)$,
where $\hat{a}$ is the annihilation operator for the cavity microwave mode. The phase shift $e^{-i\varphi_\alpha}$ accounts for the fact that this protocol squeezes along the phase space axes defined by the phase of the input coherent state. Anything less than $\Delta p^2 = \frac{1}{2}$ indicates a squeezed state. The amount of squeezing is expressed in dB, by calculating 
\be
S(\Delta p) \equiv 10{\rm log}_{10}\left(\frac{\Delta p^2}{\Delta p_{{\rm norm}}^2}\right) = 10{\rm log}_{10}\left(2\Delta p^2\right). 
\ee
In addition, we can calculate how far the state $\rho''$ deviates form a minimal uncertainty state by calculating $\Delta x \Delta p$ (where $\hat{x} = \frac{1}{\sqrt{2}}\left( \hat{a}e^{-i\varphi_\alpha} + \hat{a}^{\dagger}e^{i\varphi_\alpha}\right)$ is the conjugate observable to $\hat{p}$). Figures \ref{fig:DeltaP}  and \ref{fig:HB} show $S(\Delta p)$ and $\Delta x \Delta p$ respectively as functions of $\al$ and the number of detection events on either side of the displacement, $N$.

As can be see in figure \ref{fig:DeltaP}, the maximum amount of squeezing possible on a given input state $\ket{\al}$ increases monotonically with $\abs{\al}^2$, proportional to the power of the input pulse. Interestingly, for a given $\al$, there exists a finite $N$ that achieves a global minimum in $\Delta p^2$. One would be tempted to use this value of $N$ in the protocol to create $\rho''$; however, as can be seen in figure \ref{fig:HB}, there are other concerns.

As figure \ref{fig:HB} shows, for a given $\al$, $\Delta x \Delta p$ is not monotonic in $N$. Since we want $\rho''$ to be as close to a minimal uncertainty state as possible, while still maintaining a significantly squeezed quadrature, the optimal choice of $N$ for the protocol would be at the $\Delta x \Delta p$ local minima shown in figures \ref{fig:DeltaP} and \ref{fig:HB} by the white curve. While this does not minimize $\Delta p^2$ (and therefore maximize squeezing), it achieves a significantly squeezed state $\rho''$ that is as close to being a minimal uncertainty state as is possible, which is what we consider optimal.

\subsection{Generalized Squeezed States}
\label{sec:2B}

\begin{figure}[h]
\begin{overpic}[width=\columnwidth]
{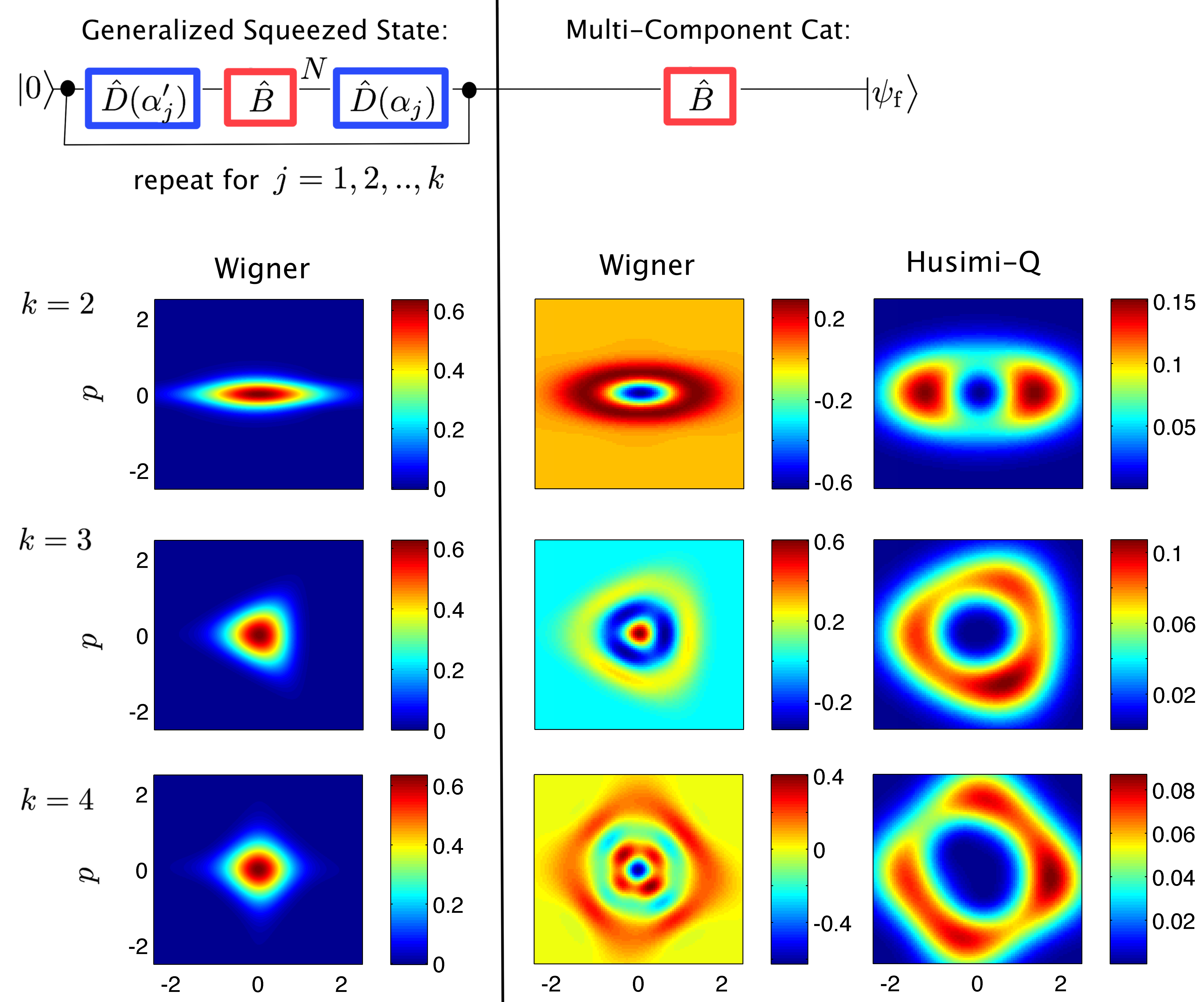}
\end{overpic}
\caption{(upper) a schematic description of the state preparation protocol. (lower) Wigner representations (first and second columns) and Husmi Q representations (third column) of the $k=2$, $k=3$, and $k=4$ generalized squeezed vacuum states (first column) and squeezed k-component Schr\"{o}dinger cat states (second and third columns).    
Appendix \ref{ap:Numerics} contains a table of the fidelities, success probabilities, and $\alpha$, $\delta$, and $N$ values for these states, as well as the fit parameters $z$ and $\beta$.}
\label{fig:Protocol}
\end{figure}

Other nonclassical states of microwave radiation having $\varphi_k\equiv 2\pi/k$ rotational symmetry in phase space result from generalizing this procedure (see figure \ref{fig:Protocol}). The squeezed state protocol discussed previously involves detecting the cavity state while centred at two points on a line through the origin of phase space, i.e., the case $k=2$.  To generalize this protocol, we detect $N$ photons at $k$ positions equally spaced around a circle of radius $|\alpha|$ in phase space, and take the first position on the positive real axis for simplicity.

There are $k$ steps to this generalized protocol, and for $j=0,1,..k-1$, the $j$th step is to: 1) displace by $\alpha_{j}$ so that the cavity state is centred around $|\alpha| e^{ij\varphi_k}$, 2) detect $N$ photons, and 3) displace by $\al_{j}'$, such that the cavity state is centred around the origin, where the coherent displacements parameters are 
\begin{align}
&\alpha_j =  |\alpha| e^{ij\varphi_k} \\
&\alpha_j'=- \delta_j e^{ij\varphi_k} \\
&\delta_j \approx \sqrt{n_{j}}.
\end{align}
The amplitude of the displacement to the origin, $\delta_j$, is approximately the square root of the photon number left in the cavity after $N$ detections, with a small correction accounting for the asymmetry of the intermediate states of the protocol.

Finally, after detecting $N$ photons at all k positions, we obtain the generalized squeezed state with $k$-fold symmetry, $\rho_k$.  The detection stage at each position transforms $\rho_{j1}$ to $\rho_{j2}$ according to
\be
{\rho}_{j2}\equiv\frac{\hat{B}^N\rho_{j1}\hat{B}^{\dagger  N}}{P_N^{j}},
\ee
where $P_N^{j}$ is the probability of detecting $N$ photons from the state $\rho_{j1}$. The entire
protocol will complete with success probability
\be
\label{eqn:SP}
{\rm Prob}({\rm success}) = \prod_{j=0}^{k-1} P_N^j,
\ee
which can be made close to unity (see appendix \ref{ap:Prob}). 
  
As mentioned previously, the case $k=2$ ($\varphi_k = \pi$) corresponds to the creation of the vacuum with squeezed quadratures.  We find high (here and henceforth meaning above 99\%) overlap with 
\be
\label{eqn:SqVac}
\ket{\Psi_2} = S(z)\ket{0} = e^{-\frac{1}{2}\left( z(\hat{a}^{\dagger})^2 - z^*(\hat{a})^2\right)}\ket{0}
\ee
by numerically searching over $N$ and $z$,
where $S(z)$ is the ordinary squeezing operator with complex squeezing parameter $z$.
In general, for $k \ge 2$, the states created by this protocol have high overlap with the analytic states
\be
\ket{\Psi_k} = S^{(k)}(z) \ket{0} = e^{-\frac{1}{2}\left( z(\hat{a}^{\dagger})^k - z^*(\hat{a})^k\right)}\ket{0},
\ee
where $S^{(k)}(z)$ is called a generalized squeezing operator with complex parameter $z$ \cite{Braunstein:1987kx,Braunstein:1990vn}. The first three operators of this class are 
\bea
\nonumber &&S^{(0)}(z)  = e^{-{\rm Im}[z]}\mathbb{I}, \  S^{(1)}(z)  = D\left(-\frac{z}{2}\right) \\
&&{\rm and }\ S^{(2)}(z)  = S(z), 
\eea
where $S(z)$ is the squeezing operator of equation (\ref{eqn:SqVac}).

Consider the final state after displacing to the origin $\rho_k$.  To find $z$, we impose the condition that $\rho_k$ and $\ket{\Psi_k}$ have the same average photon number by setting
\be 
\bra{\Psi_k}\hat{a}^\dagger \hat{a}\ket{\Psi_k} = \big<\hat{a}^\dagger \hat{a}\big>_{\rho_k}.
\ee
This results in excellent fidelity between the states, quantified by
\be
{\rm F}\left[\rho_k\right] \equiv {\rm Tr}\left\{\rho_k\ket{\Psi_k}\bra{\Psi_k}\right\}.
\label{eqn:Fidelity}
\ee
In fact, for $k = 2,3,4$ the fidelity is greater than 99\% in each case for various values of $\abs{\al}$. The Wigner functions of these states are shown in the left column of figure \ref{fig:Protocol}.

To verify that these states are indeed nonclassical, a suitable nonclassicality witness can be used. For this purpose, we use the entanglement potential of \cite{Asboth:2005fk}, where a nonzero entanglement potential indicates that a state is nonclassical. Indeed, as can be calculated numerically, the states created by the generalized protocol all have nonzero entanglement potential (see appendix \ref{ap:NC}).

Since the first displacement stage of each step ensures the average photon number of each state $\rho_{j1}$ is on the order of $\abs{\al}^2$, each $P_N^j$ can be quite large, and as a result, the generalized protocol can have a significant probability of success. For example, for the modest $\abs{\al}$ of the states shown in figure \ref{fig:Protocol}, the success probabilities are all greater than 99\%. Furthermore, the success probability will grow monotonically with $\abs{\al}$, and so can be increased by increasing the initial input state power. 

\subsection{Squeezed Multi-Component Cat States}

It is known in quantum optics  \cite{Dakna:1997fk} that subtracting a photon from a squeezed vacuum produces an odd  Schr\"{o}dinger cat state.   This concept can be generalized, such that  subtracting a photon from a generalized squeezed vacuum of $k$-fold symmetry (the output state $\rho_k$ of our protocol in \ref{sec:2B}) leaves a squeezed, $k$-component Schr\"{o}dinger cat state in the cavity.   
We note that while having some similarity to the optical setting, our protocol does not require a beam splitter or a photon number resolving detector, instead, it simply requires the subtraction of one more photon from the squeezed states created by the protocol described in the previous section.

If we remove one more photon from the final state $\rho_k$, the resulting state has the form
\be
\rho'_k = \frac{\hat{B}\rho_{k}\hat{B}}{P_1^{k}}^\dagger,
\ee
where $P_1^{k}$ is the probability of a single photon being detected from state $\rho_k$. This procedure produces states of high overlap with
\be
\label{eqn:gensqcat}
\ket{\Psi'_k} = S^{(k)}(z)\sum_{j = 0}^{k-1}e^{ij(\varphi_k+\pi)}\ket{\beta e^{ij(\varphi_k+\pi)}},
\ee
where now both $z \in \mathbb{C}$ and $\beta \in \mathbb{R}$ must be found numerically. These states all have non-zero entanglement potential, and are therefore nonclassical states of microwave radiation (see appendix \ref{ap:NC}).

For $k=2$, this additional detection will create a state very close to a squeezed odd Schr\"{o}dinger cat state
\be
\ket{\Psi'_{2}} = S^{(2)}(z) \left( \ket{\beta} - \ket{-\beta} \right).
\label{eqn:SCat}
\ee
For $k = 3$ and $k=4$, we have
\begin{align}
&\ket{\Psi'_{3}} = S^{(3)}(z) \left(e^{-i\frac{\pi}{3}}\small|\beta e^{-i\frac{\pi}{3}}\small\rangle + e^{i\frac{\pi}{3}}\small|\beta e^{i\frac{\pi}{3}}\small\rangle - \ket{-\beta}\right) \label{eqn:SVoo} \\
&\ket{\Psi'_{4}} = S^{(4)}(z) \left( \ket{\beta} + i\ket{i\beta} - \ket{-\beta} - i\ket{-i\beta}\right).
\label{eqn:SZen}
\end{align}
The $k=3$ state is the state formed when the operator $S^{(3)}(z)$ is applied to a voodoo cat state \cite{Hofheinz:2009qf}. The $k=4$ state is the operator $S^{(4)}(z)$ applied to a coherent superposition of four out of phase coherent states, known as a compass state, which is known to have favourable decoherence properties \cite{Zurek:2001fr,Kirchmair:2013uq}. The Wigner and Husimi Q functions of these squeezed multi-component cat states are shown in the middle and right columns of figure \ref{fig:Protocol} respectively, all of which have greater than $94\%$ fidelity with (\ref{eqn:SCat}),  (\ref{eqn:SVoo}), and (\ref{eqn:SZen}).  

We have plotted both representations of these states as they highlight distinct information about the state. The Q function emphasizes the cat-like properties of the final state, while the Wigner function makes apparent the similarity between the final state and a $k-1$ photon Fock state, squeezed to the same order in $k$. The highly nonclassical nature of the state is also made evident by the large negative region of its Wigner function.

The probability of successful generation of these multi-component cat states is
\be
\label{eqn:SCP}
{\rm Prob}({\rm success})' = P_1^k \prod_{j=0}^{k-1} P_N^j.
\ee
Unfortunately, the $P_1^k$ are often very small, and for the squeezed multi-component cat states shown in figure \ref{fig:Protocol} this results in a much lower success probability than the generalized squeezed vacuum.  Optimization of this success probability is discussed in appendices \ref{ap:Prob} and \ref{ap:Numerics}.

\section{Discussion}

\subsection{Experimental Implementation}
In regards to experimental implementation of this protocol with JPMs, superconducting microwave resonators are currently fabricated  with Q-factors approaching $10^7$, which in the microwave regime will lead to cavity lifetimes on the order of $10^5$ns \cite{Megrant:2012uq}. Thus, when JPMs with short $T_2$ are used \cite{Govia:2012uq}, we can conservatively expect that as many as $10^2$ to $10^3$ measurements can be performed in the lifetime of the cavity.  Practically, space requirements on a chip require a fast reset strategy for the JPMs, which is currently being developed \cite{McDermott12}.

For a realistic implementation of this protocol via JPMs, one must also consider the possibility of energy dissipation and dark counts in the JPMs. The behaviour of a JPM under such conditions has previously been discussed in \cite{Govia:2012uq,Poudel:2012uq}, and as such, we will highlight only the key point here. Both energy relaxation in the detector, and dark counts can be treated on the same footing. Energy relaxation corresponds to an unregistered measurement, and dark counts to false positives. One can thus adjust the number of measurements according to these rates to approach the desired number of $\hat{B}$-applications. 

Moreover, as is illustrated in figure \ref{fig:DeltaP}, the asymmetric squeezing performed at the $j$'th step is a smooth and slowly varying function of $N$, and as such is only minimally affected by a small change in $N$. Thus, the protocol is generally robust to the effects of energy relaxation and dark counts. Only the final detection used to generate squeezed multi-component cat states is sensitive to small perturbations in $N$, with an odd number of detections producing states of the form seen in equation (\ref{eqn:gensqcat}). 

Photon lifetimes are in general a limiting factor for nonclassical states, but these have been reported to be very long in circuit QED (on the order of 100 $\mu$s in \cite{Kirchmair:2013uq}), and given the fast measurement times of our protocol, the cavity lifetime will affect our protocol less than others. In addition, some of the nonclassical states created by our protocol are robust to photon loss. For example, the even $k$-th order squeezed states are robust to single photon loss \cite{Leghtas:2013fk}.

\subsection{Applications}

The generalized squeezed states have applications in continuous variable quantum computing \cite{Lloyd:1999fk}. Our protocol provides a simple way to obtain the necessary single mode squeezing and nonlinearity: by applying the generalized squeezed state protocols for $k=2$ and $k\ge3$ respectively. In particular, implementing the nonlinearity is often the technological bottleneck, as it cannot be created using linear optics alone \cite{Bartlett:2002uq}, and our protocol has the potential to be more efficient and require less technological overhead than most known methods to implement the nonlinearity \cite{Bartlett:2002uq, Milburn:1983kx}.

The squeezed multi-component Schr\"{o}dinger cat states (SMCS) have applications in metrology, in particular for phase estimation using a Mach-Zehnder interferometer set up. It has recently been shown that cat states can be used in combination with linear optics to create Entangled Coherent states (ECS), and that phase estimation using these ECS outperforms that using NOON states \cite{Joo:2011fk}. The performance difference is especially significant at low photon number, and/or when photon loss is considered. It is an emerging and active area of research to see if an improvement in phase estimation can be gained by using SMCS in place of cat states in this scheme.

\subsection{Conclusions}

In conclusion, we have shown how a combination of strongly coupled photon counting and coherent displacement can be used to create nonclassical states of radiation with high probability. This protocol can be realized in circuit QED using Josephson photomultipliers. 

We acknowledge instructive discussions with F.W. Strauch and R. McDermott. Research supported by DARPA within the QuEST program and NSERC Discovery grants. LCGG was supported by the Ontario Graduate Scholarship Program.

\section{Appendices}

\appendix
\section{Nonclassicality}
\label{ap:NC}

\begin{figure}[h!]
\includegraphics[width = \columnwidth]{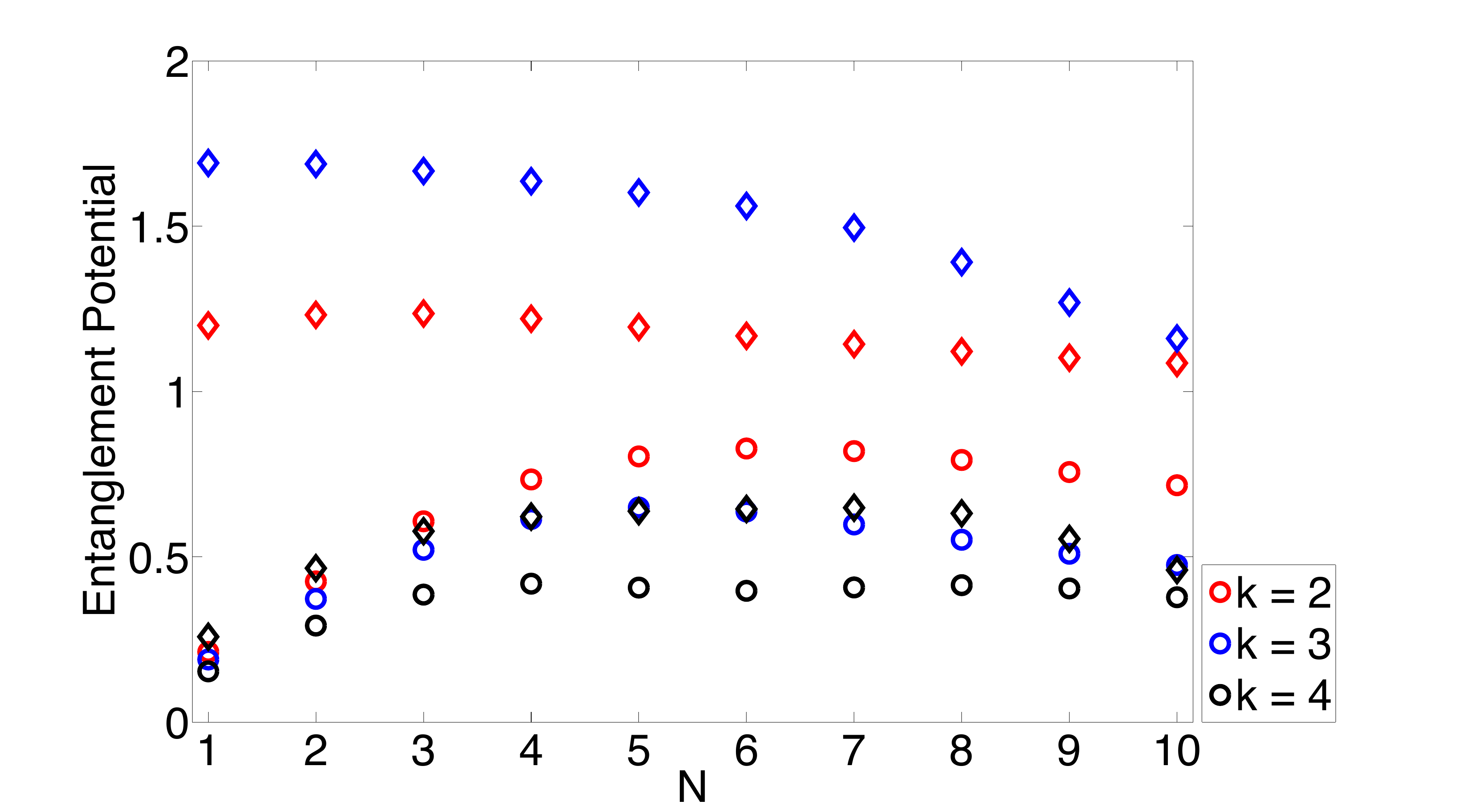}
\caption[Entanglement Potential]{This figure shows entanglement potential (EP) of the states created by the generalized protocol (for $k = 2,3,4$) as a function of the number of detections, $N$, performed at each step in the protocol. The generalized squeezed states are the circles, and the diamonds correspond to the squeezed multi-component cat states. These states all have $\abs{\al} = 3$. }
\label{fig:EPPlot}
\end{figure}

The entanglement potential is both a nonclassicality witness and measure, and as such can be used to examine how the nonclassicality changes with the number of detections at each step of the generalized protocol. It is defined as
\be
{\rm EP}[\rho] \equiv {\rm log}_2\abs{\abs{\sigma_{\rho}^{T_A}}}_1,
\ee
where $\sigma_{\rho} = U_{\rm BS}\left( \rho \otimes \ket{0}\bra{0}\right)U_{\rm BS}^{\dagger}$, for $U_{\rm BS}$ the unitary transformation of a 50:50 beam splitter, and $\sigma_{\rho}^{T_A}$ is the partial transpose of $\sigma_{\rho}$ \cite{Vidal:2002uq}. Figure \ref{fig:EPPlot} shows the entanglement potential of the generalized squeezed states and the multi-component cat states (for $k = 2, 3, 4$) as a function of the number of detections at each step. As can be seen, for the squeezed state ($k=2$), the maximum value of EP occurs around the optimal value of $N$ determined in the main text.

\section{Success Probability vs. Fidelity}
\label{ap:Prob}

\begin{figure}[h!]
\subfigure{
\label{fig:PN}
\begin{overpic}[width=\columnwidth]
{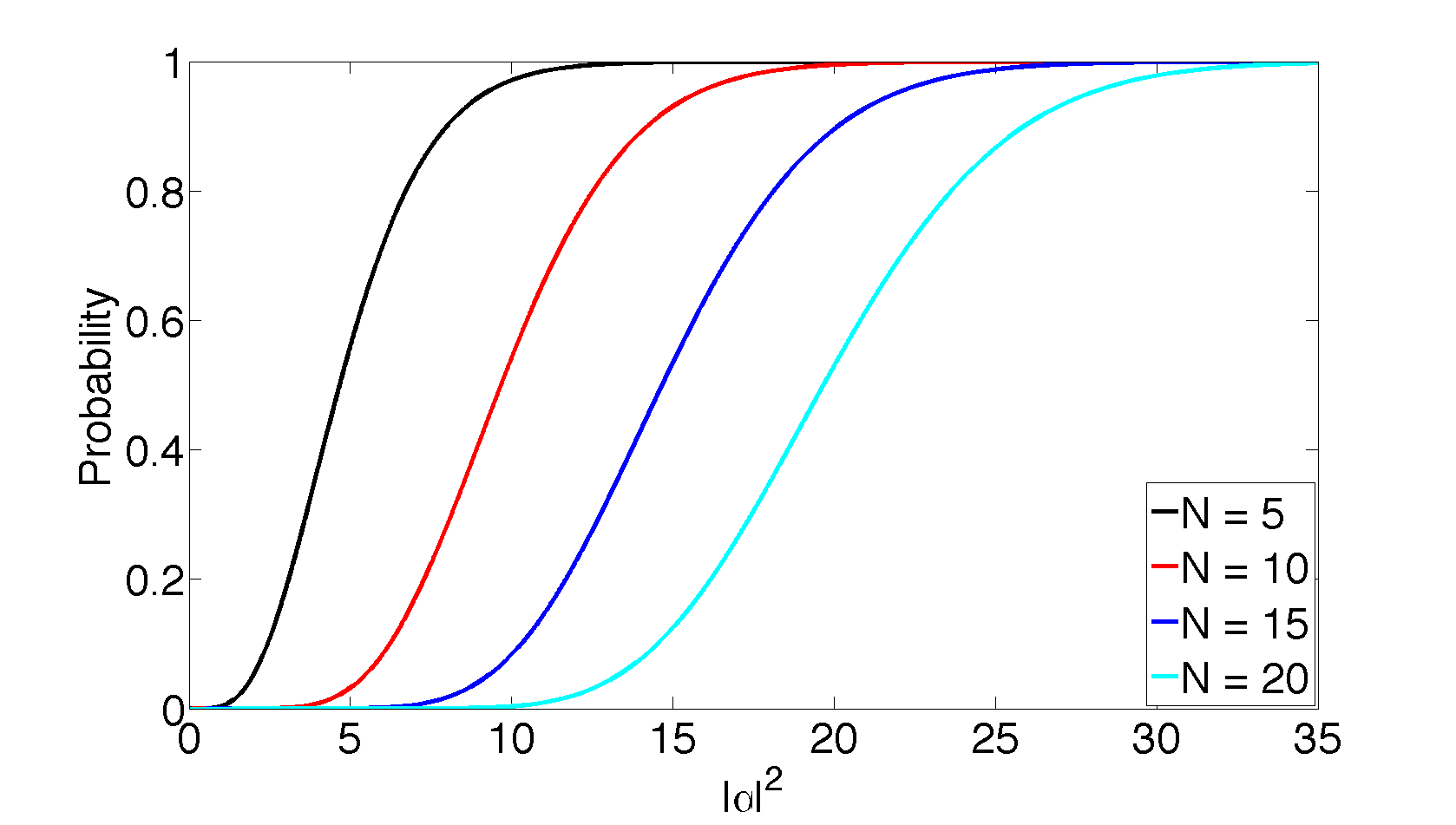}
\put(40,115){{\color{black} \bf (a)}}
\end{overpic}}
\subfigure{
\label{fig:PNMod}
\begin{overpic}[width=\columnwidth]
{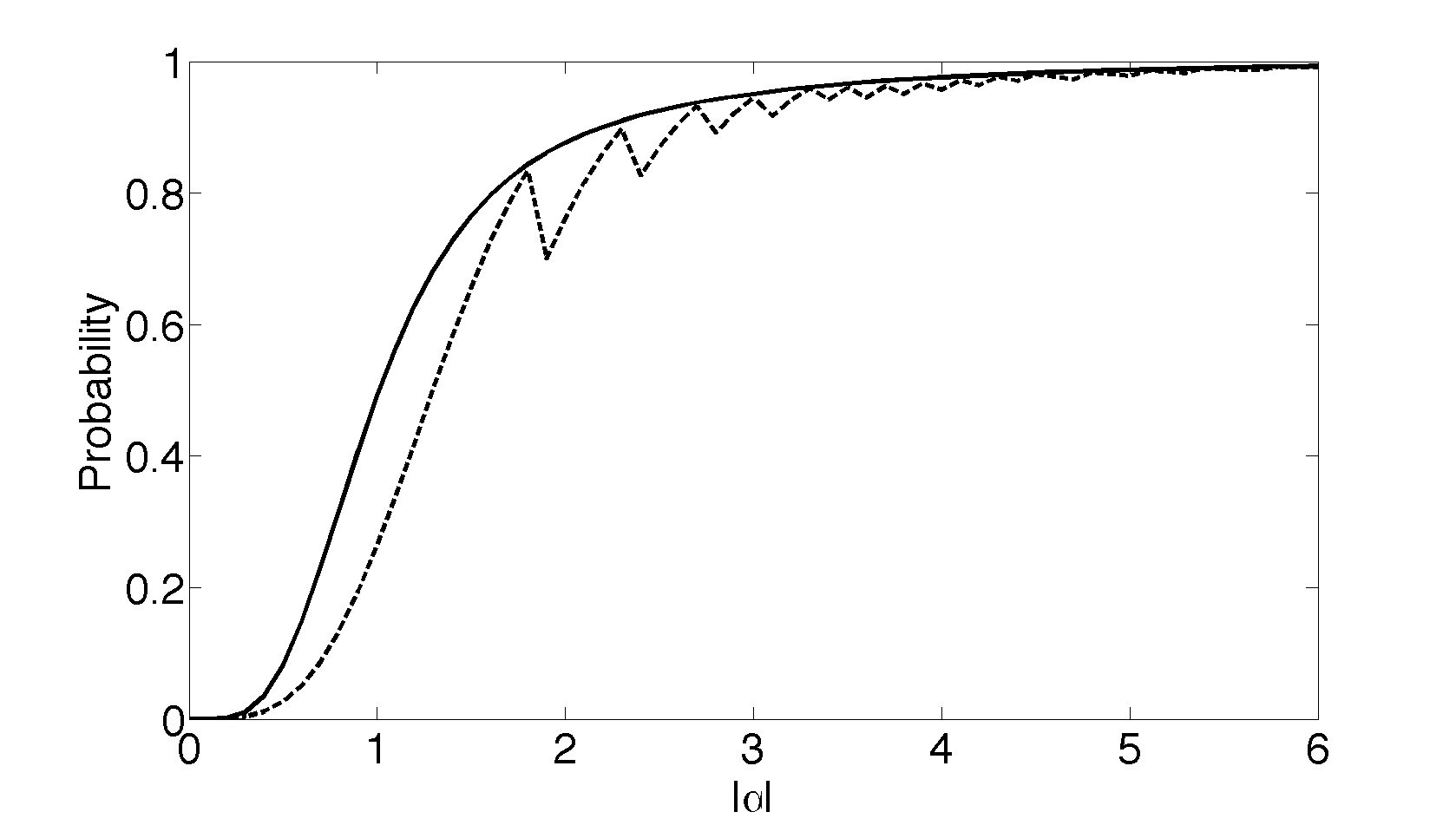}
\put(40,115){{\color{black} \bf (b)}}
\end{overpic}}
\caption{The success probability of the first step of the protocol as a function of input coherent state power. In (a) for several values of fixed $N$, and in (b) for N assumed to vary quadratically as $q(\abs{\al})$. The solid line in (b) is for $N$ taking real values, and the dashed line for $N$ taking only integer values.}
\end{figure}

Due to the noncommutivity of the subtraction operator and coherent displacement, finding a simple closed form analytic solution for the success probability of our protocols (equations (\ref{eqn:SP}) and (\ref{eqn:SCP})) may not be possible. However, we can examine the behaviour of $P_N$ of equation (\ref{eqn:ProbN}) to understand how the total detection probability scales with $\abs{\al}^2$. As can be seen in figure \ref{fig:PN}, $P_N$ rapidly approaches unity for $\abs{\al}^2 > N$. As a result of this, it is possible to achieve very high success probabilities for the generalized squeezed state protocol. For the squeezed multi-component cat states, the limiting factor remains $P_1^k$, which can be quite small. 

In addition to success probability, one also wishes to maximize the fidelity with the target analytic states. In the $k=2$ case, we find that this is achieved when, for a given $\abs{\al}$, $N$ lies nearly along the minimal uncertainty curve of figure \ref{fig:HB}. This curve is well approximated by $q(\abs{\al}) = a\abs{\al}^2 + b\abs{\al} + c$, where $a,b,c \in \mathbb{R}$ can be found numerically. Since $N$ must be an integer, we set 
\be
N = \lceil q(\abs{\al})\rceil,
\ee
where $\lceil * \rceil$ rounds up to the nearest integer. It is worth examining what effect this has on the success probability of $P_N$, now defined by
\be
\label{eqn:PNMod}
P_N= 1 - \frac{\Gamma(\lceil q(\abs{\al})\rceil,\abs{\al}^2)}{\Gamma(\lceil q(\abs{\al})\rceil)}.
\ee
This is plotted in figure \ref{fig:PNMod}, along with the continuous version of equation (\ref{eqn:PNMod}), where $N$ is allowed to take non-integer values.
As the figures show, the success probability approaches unity with increasing $\abs{\al}$. It is therefore possible to maximize both fidelity and success probability in the $k=2$ case, for both the squeezed state and Schr\"{o}dinger cat state.

For $k>2$ there is no known analog of the minimal uncertainty curve of figure \ref{fig:HB}, however, numerical results have shown that it is possible to achieve high fidelity and success probability using $N$ and $\abs{\al}$ similar to that of the $k=2$ case, i.e. near to the minimal uncertainty curve.

\section{Numerical Results}
\label{ap:Numerics}
In figure \ref{fig:Protocol}, we have given examples of a generalized squeezed  vacuum and a squeezed/multi-component cat, for $k=2,3,4$.   Each of these examples is found by a numerical fit over the parameters in the protocol:  the field amplitude $\alpha$, the number of photon subtractions at each position $N$, and the small corrections for photon loss, $\delta_1$, .., $\delta_k$.   For each of the examples in Fig. \ref{fig:Protocol}, we have locally minimized the error function
\begin{equation}
\epsilon=1-F[\rho_k,\rho_k^{\rm t}]
\label{eqn:error}
\end{equation}
over the set of parameters ${\cal P}=\{x_i\}$ necessary to define $\rho_k$ -- the final state produced in our protocol -- and the target state $\rho_{k}^{\rm t}=|\Psi_k\big>\big<\Psi_k|$.   
The fidelity $F$ is as defined in equation (\ref{eqn:Fidelity}).  We have given examples only for $N=16$ ($N=6$ in the case of the 4-component cat state) to reduce this parameter space.  Although this choice is somewhat arbitrary, it generally predicts reasonably small error for $\abs{\alpha}\in[3,10]$, an experimentally accessible range.
\begin{widetext}

\begin{figure}[h!] 
   \centering
   \footnotesize
  \begin{tabular}{ l || l }
  {\it generalized squeezed vacuums} & {\it squeezed multi-component cats}  \\
  \hline
  \begin{tabular}{l || l  l l l l}
  k=2 & $\alpha=5.7389$&&&\\
 &$\delta_1=1.6786$&$\delta_2=1.7227$ &&\\
    &$z=0.7063$&&&\\
 &\multicolumn{4}{l}{$P=0.9951$ \ $\epsilon^{(2)}=1.7\times 10{-3}$  } \\ 
  \end{tabular}
  &
    \begin{tabular}{l l l l}
$\alpha=3.8480$&&&\\
$\delta_1=1.9783$&$\delta_2=1.5670$&&\\
\multicolumn{4}{l}{$z=-3.8667$\ $\beta=2.00$}\\
$P=0.2478$ & $\epsilon^{(2)}=2.0\times10^{-3}$ && \\ 
  \end{tabular}\\
  \hline
  \begin{tabular}{l || r  l l l l l}
  k=3 &  $\alpha=6.6999$&&&&&\\
  &$\delta_1=1.1782$&$\delta_2=1.1536$&$\delta_3=1.1532$&&&\\
  &$z=-0.185$&&&&&\\
 &\multicolumn{5}{l}{$P=1.0000$ \ $\epsilon^{(3)}=4.5\times10^{-4}$  } \\ 
  \end{tabular}
  &
    \begin{tabular}{l l l l l}
$\alpha=7.5000$&&&&\\
$\delta_1=1.582$&$\delta_2=1.1522$&$\delta_3=1.1482$&&\\
\multicolumn{5}{l}{$z=4.4\times 10^{-2}$ \ $\beta=6.0\times 10^{-5}$}\\
$P=0.0037$ & \multicolumn{2}{l}{$\epsilon^{(3)}=4.4\times10^{-3}$} && \\ 
  \end{tabular}\\
  \hline
  \begin{tabular}{l || r  l l l l l l}
  k=4 & $\alpha=5.9999$&&&&&&\\
  &$\delta_1=1.2138$&$\delta_2=1.2166$&$\delta_3=1.2047$&$\delta_4=1.2193$&&&\\
  &$z=1.4776$&&&&&&\\
 &\multicolumn{6}{l}{$P=0.9999$ \ $\epsilon^{(4)}=6.1 \times10^{-3}$  } \\ 
  \end{tabular}
  &
    \begin{tabular}{l l l l l l}
$\alpha=3.5500$&&&&\\
$\delta_1=0.9451$&$\delta_2=0.9518$&$\delta_3=0.9393$&$\delta_4=0.9401$&\\
\multicolumn{6}{l}{$z=0.0014$ \  $\beta=1.56$}\\
$P=0.0025$ &\multicolumn{2}{l}{ $\epsilon^{(4)}=5.6\times10^{-2}$  } &&\\ 
  \end{tabular}\\
  \hline
\end{tabular}
   \caption{Protocol parameters $\alpha$ and  $\{\delta_j\}$, target state parameter(s) $z$ (and $\beta$), error $\epsilon^{k}$, and corresponding success probability $P$ for each of the examples given in Fig. \ref{fig:Protocol}.  $N=16$ in all cases except for the 4-component cat, in which case $N=6$.  }
\label{fig:Parameters}
\end{figure}

\end{widetext}  

In the case of the generalized squeezed vacuum states, we constrain the squeezing parameter $z$, all that is necessary to specify the target state,  so that 
\begin{equation}
\big<\hat{n}\big>_{\rho_{k}}=\big<\hat{n}\big>_{\rho_k^{\rm t}}.
\label{eqn:constraint}
\end{equation}
There remains $k+1$ unconstrained parameters determining the error function:
\begin{equation}
\epsilon^k\equiv\epsilon^k(\alpha,\delta_1,..,\delta_k).
\label{eqn:parameters}
\end{equation}
In the case of multi-component/squeezed cats, two parameters specify $\rho_k^{\rm t}$, $z$ and the cat-state amplitude $\beta$.  In this case, $\beta$ is constrained according to equation (\ref{eqn:constraint}), and we add $z$ to the set of parameters over which we minimize the error:
\begin{equation}
\epsilon^k\equiv\epsilon^k(z,\alpha,\delta_1,..,\delta_k).
\label{eqn:parameters}
\end{equation}
Figure \ref{fig:Parameters} contains the parameter values, $\epsilon^{k}$, and success probabilities $P$ for each of the cases shown in figure \ref{fig:Protocol}. 

Note that the error function equation (\ref{eqn:error}) is multimodal, so we have not likely found a global maximum in state fidelity.   Generally, the detection probability is also multimodal when the integer nature of $N$ is considered, e.g. see dotted curve in figure \ref{fig:PNMod}, so the detection probabilities listed here are not likely maximal and depend sensitively on the choice of $N$.  While our choice $N=16$ is well-suited for $k=2$, other values would likely optimize the cat-state detection probabilities for different $k$.  We leave the optimal value of $N$ for a given $k$ as an open question.   

\bibliography{NCLbib}

\end{document}